\documentstyle[prd,aps,epsfig]{revtex}

\newcommand{\eq}{\begin{equation}}
\newcommand{\beqn}{\begin{eqnarray}}
\newcommand{\en}{\end{equation}}
\newcommand{\eeqn}{\end{eqnarray}}
\def\part{\partial}

\begin{document}
\draft
\preprint{UIUC-THC-98/6}
\title{Living With Lambda}
\author{J.D. Cohn}
\address{Departments of Physics and Astronomy,
University of Illinois \\
Urbana, IL 61801 \\
{\rm jdc@uiuc.edu} \\}
\date{July 1998}
\maketitle
\begin{abstract}
This is a short pedagogical introduction to
the consequences of a nonzero cosmological constant $\Lambda$ in physical
cosmology.
\end{abstract}
\bigskip

The cosmological constant is an
energy associated with the vacuum, that is, with ``empty space''.
The possibility of a nonzero cosmological constant $\Lambda$
has been entertained
several times in the past for theoretical and observational reasons
(early work includes e.g. Einstein 1917,
Petrosian, Salpeter \& Szekeres 1967,
Gunn \& Tinsley 1975, a popularized description of the history
is found in Goldsmith 1997).
Recent supernovae results (Perlmutter et al 1998, Riess et al 1998)
have made a strong case for
a nonzero and possibly quite large cosmological
constant.  Their results have
encouraged increased interest in the properties
of a universe with nonzero cosmological constant.
Several other observations of various cosmological phenomena
are also planned or underway which will
further constrain the range of allowed
values for the cosmological constant.
Given the expected quality and quantity of
upcoming data, there is reason to believe that we will
know soon whether or not we need to learn to
``live with lambda''.

The purpose of this review is to provide a short pedagogical
introduction to the consequences of a nonzero cosmological constant.
Basic terms are defined in section one, where the equations for
the time evolution of the scale factor of the universe (defined below) are
given.  Section two indicates the current theoretically
expected values of the cosmological constant, introducing the
theoretical ``cosmological constant problem.''  Some suggestions
to explain a cosmological constant consistent
with current measurements are listed.
In section three, the time evolution of the scale factor of the
universe (from section one) is used to show how the age of
the universe, the path length travelled by light, and other
properties depending on the spacetime geometry
vary when the cosmological constant is present.
Section four outlines some effects of a nonzero cosmological
constant on structure formation.
Section five summarizes how some recent and upcoming
measurements may constrain $\Lambda$.  
Several observations (including the supernovae) are 
described which have
provided constraints or show promise for the future.
These observations are improving rapidly.
The current theoretical explanations for a nonzero cosmological constant
consistent with the data, some of which are listed in section two,
are not compelling.  Section six briefly describes
some suggested theoretical alternatives to a nonzero cosmological constant.
Section seven contains a description of the future of the universe if
the cosmological constant is nonzero and then summarizes.

Earlier reviews, in particular
the one by Carroll, Press and Turner (1992,
hereafter denoted by CPT)
are highly recommended for some of the in depth results and more
references, as well as
the books by Kolb \& Turner (1990), Peebles (1993) and
Padmanabhan (1993) for the basic cosmology.
The referencing is indicative rather than comprehensive, for
more extensive referencing consult the more in depth reviews,
textbooks and articles cited.

\section{Introduction}

The cosmological constant is a constant energy density associated with
``empty space.''  Its presence affects the properties of
spacetime (specified by a metric) and matter (stress energy).
The metric of spacetime and the stress energy tensor
of matter are related via Einstein's equations.

Thus a first step in studying the effects of the cosmological
constant is to specify the metric and stress energy tensor and
then to relate them via Einstein's equations.
For a spatially homogeneous spacetime
the metric can be written as
\eq
ds^2 =  c^2 dt^2 - R^2(t)\{d\chi^2 + r^2(\chi) [d \theta^2 +
\sin^2 \theta d \phi^2 ]\} \; .
\label{metric}
\en
In the above,
the expansion or contraction of space with time is given by $R(t)$,
and $d \chi = dr/(1-kr^2)^{1/2}$, with $k=(-1,0,1)$ for an open,
flat, or closed universe respectively.
Thus, using the notation of CPT,
\eq
r(\chi) = {\rm sinn} \; \chi =
\left\{
\begin{array}{llll}
\sin \chi \; \; \; \; &{\rm if}\; \; \; & k =1 \; \; \; \;
&{\rm closed} \\
\nonumber \chi &{\rm if}& k=0 & {\rm flat}\\
\nonumber \sinh \chi &{\rm if}&k = -1 &  {\rm open}
\end{array}
\right.
\label{rchi}
\en
Hereon, the speed of light, $c$ will be set to one,
so that a mass $m$ and the corresponding energy $mc^2$ are both
denoted by $m$.

The matter background will be assumed to be isotropic and homogeneous,
described by a stress tensor $T^\mu_\nu$ 
of a fluid in its rest frame.
Thus in the $c=1$ notation used here
$T^\mu_\nu = {\rm diag} (\rho,-p,-p,-p)$, where $\rho$ is
energy/matter density and $p$ is pressure.     
The Einstein equations are
(in terms of the Ricci tensor and scalar curvature
which will not be needed subsequently)
\eq
({\cal R}_{\mu \nu} - \frac{1}{2} g_{\mu \nu} {\cal R}) =
8 \pi G T_{\mu \nu} \; .
\en
For a spacetime with
metric and matter given above, and a
cosmological constant $\Lambda$, these equations of motion yield
\eq
\left(\frac{\dot{R}}{R}\right)^2 = H^2 
= \frac{8 \pi G \rho}{3} - \frac{k}{R^2} + \frac{\Lambda}{3} \; .
\label{eom}
\en
Here, $\dot{R} = dR/dt$, etc.
In addition the acceleration of $R(t)$ obeys
\eq
\frac{\ddot{R}}{R} = -\frac{4 \pi G}{3}(\rho + 3 p) + \frac{\Lambda}{3}
\label{accel}
\en
where $p$ is pressure in the matter stress tensor above.

Write $H_0$ for the numerical value of $H$ today, the Hubble constant,
$H_0 = 100 \, h$ km/s/Mpc  and $R_0$ for the
value of the scale factor today.
Then define
\eq
\begin{array}{l}
{\displaystyle \Omega_k = \frac{-k}{R_0^2 H_0^2}} \\
{\displaystyle \Omega_\rho = \frac{8 \pi G }{3 H_0^2} \rho_0} \\
{\displaystyle \Omega_\Lambda = \frac{\Lambda}{3 H_0^2} } \; ,
\end{array}
\en
constants corresponding to the values of these ratios today.
The equation of motion, equation (\ref{eom}), implies
\eq
1 = \Omega_k + \Omega_\rho + \Omega_\Lambda \; .
\en

The constants $\Omega_k,\Omega_\rho,\Omega_\Lambda$ can be used to
rewrite equation (\ref{eom}) at a general time.
Defining the scale factor $a(t) = R(t)/R_0$,  $a=1$ today, and
the equation of motion (\ref{eom}) becomes
\eq
\left(\frac{\dot{a}}{a} \right)^2 =
\frac{8 \pi G \rho}{3} - \frac{k}{a^2 R_0^2}
+ \frac{\Lambda}{3}
\label{aeom}
\en

The $a$ dependence of the terms proportional to the
curvature $k$ and $\Lambda$ can be read off directly.
The energy density $\rho$ may have
several components which scale differently as the universe
expands.  Nonrelativistic (``cold'') matter
scales as the volume of the universe,
\eq
\Omega_{NR}(a) = \Omega_{NR} (R/R_0)^{-3} = \Omega_{NR} a^{-3}
\en
while relativistic matter, such as radiation, redshifts,
\eq
\Omega_{R}(a) = \Omega_R a^{-4} \; .
\en
In addition to these two familiar possibilities,
other sorts of matter are possible
which scale as other powers of $a$ and
are discussed in section six.
Thus generally the energy density term $\Omega_\rho$ has several parts:
\eq
\Omega_\rho(a) = \Omega_{NR}(a) + \Omega_{R}(a) +  \cdots
\label{omegagen}
\en

As the universe expands,
the energy density of the relativistic
matter decreases faster than that of the non relativistic matter.
In the standard cosmological scenario, earlier on the 
universe was radiation
dominated, that is the relativistic energy density was the
largest component of the matter energy density.
Now non-relativistic matter makes up a larger
fraction of the energy density (matter domination).
For example, $\Omega_{R, \, {\rm CMB}}$, the contribution from the
cosmic microwave background (CMB)
photons today, is about $10^{-5}$, and is becoming increasingly
negligible as $a$ increases.

Including only relativistic and nonrelativistic matter for now,
equation \ref{aeom} above can be rewritten as
\eq
\left(\frac{\dot{a}}{a}\right)^2 =
H_0^2(
\frac{\Omega_{NR}}{a^{3}} + \frac{\Omega_R}{a^4} +
\frac{\Omega_k}{a^{2}} + \Omega_\Lambda),
\en
with square root
\eq
\frac{1}{a }\frac{da}{dt}  =
H_0 \left(
\frac{\Omega_{NR}}{a^3} + \frac{\Omega_R}{a^4} +\frac{\Omega_k}{a^2} +
\Omega_\Lambda\right)^{1/2}
= H_0 E(z = a^{-1} - 1) \; .
\label{Ez}
\en
The generalization of $\dot{a}/a$ and hence
$E(z)$ to more general $\Omega_\rho$ is immediate once
the dependence on the scale factor $a$ is known.

The acceleration equation can be
written in terms of $\Omega_{NR}$, etc.,
as well, by including the different
equations of state $p_a = w_a \rho_a$ for
 the different sorts of matter.
If only relativistic and nonrelativistic 
matter is present, equation
\ref{accel} becomes
\eq
\frac{\ddot{a}}{a} = -H_0^2 (\frac{\Omega_{NR}}{2} a^{-3}
+\Omega_R a^{-4} - \Omega_\Lambda)
\en

Equation \ref{Ez} allows one to read off the effect of various
possible combinations of relativistic and nonrelativistic matter,
curvature and cosmological constant
on the history of the universe.  For example, one possibility
(depending on magnitudes and signs) is that
relativistic matter dominates for $a$ small, then
nonrelativistic matter, then curvature and
finally the cosmological constant.

The behavior of the scale factor $a(t)$ for different
$\Omega_\rho, \Omega_\Lambda$ are given in
Felten \& Isaacman (1986).  The value of $\Omega_k$ determines whether
the universe is spatially closed, open or flat, while models which
expand forever or recollapse are determined by whether
$\dot{a}$ changes sign, passing through zero.
One can see from equation (\ref{Ez})
that if  $\Omega_\Lambda$ is big enough, the universe will
expand forever, eventually expanding exponentially fast.

\section{Expected $\Lambda$}
Given that the cosmological constant may occur in the equations of
motion, one can ask what value of $\Lambda$ is expected.
High energy particle theory, which in principle could include a theory
of gravity, has no known way to determine the value
of the cosmological constant from first principles.  Many suggestions
have been made, some of which are listed later in this section.
However,
cosmologically interesting scales for $\Lambda$ are
puzzlingly small in particle theory.

In particle theory, field quantization allows a zero point energy,
the constant energy when all fields are in their ground state.
This has the same effect as a cosmological constant.
In the
absence of gravity, only the difference between zero points of different
systems can
be measured, while
the absolute value is unmeasurable.
(The difference between the vacuum energies or zero points
of different systems was demonstrated, for
example, with the Casimir (1948) effect:
Two parallel plate conductors provide boundary conditions
for the vacuum between them.   The vacuum energy depends on
these boundary conditions.   Changing the distance between
the plates changes the boundary conditions, and thus the vacuum energy,
resulting in a measurable force between the two plates.)
 However, Einstein's equations
above react to the value of the vacuum energy itself, and 
the
theoretical criteria for setting the absolute zero point are unclear.

The estimate of the vacuum energy in particle
theory is reviewed in CPT.  When fields are quantized in particle theory,
they can be considered as a set of harmonic oscillators, and these
each have an associated zero point energy.  
The total number of harmonic oscillators is determined by the 
high energy cutoff $E_{max}$ of the theory. If there is no cutoff,
that is, the naive field theory counting of degrees of
freedom  works to arbitrarily high
energies and short scales, then $E_{max}$ is infinite. 
Taking the sum of
zero point energies and dividing by the volume of space
gives the corresponding
contribution to the constant energy density,
\eq
 \rho_{vac} \sim  \frac{E_{max}^4}{\hbar^3} \; .
\en
An observed cosmological constant of order one corresponds to
$E_{max} \sim 10^{-2}-10^{-3}$eV.
In contrast, it is not expected that naive field theory works
at the scales where gravity is expected to become strong, so
at the very least,
a natural cutoff for particle theory is the scale of
gravity, $E_{max,gravity} \sim 10^{28}$ eV. 
There are also mechanisms within the context of field
theory which can produce a cutoff, lowering $E_{max}$.  For
instance, supersymmetry relates particles
of different spin.  Supersymmetry can cause
cancellations between
zero point energies of different
particles down to the scale where
supersymmetry breaks,
so that
$E_{max,susy} \sim 10^{20}$eV. 
In most cases the expected particle
physics scales for the cutoff
give $E_{max} >100 \; {\rm GeV} =  10^{11}$eV, so that
the vacuum energy density $\rho_{vac}$, 
proportional to $\Lambda$, obeys
\eq
\frac{\rho_{vac}^{obs}}{\rho_{vac}^{pred}} <
\frac{(10^{-2})^4} {(10^{11})^4}
= 10^{-52}
\en
This huge disparity between the naively expected $\Lambda$ 
and the value of  $\Lambda$ consistent with observations
is, for
particle physicists, `the cosmological constant problem.'
The relatively tiny value for a cosmologically
relevant $\Lambda$ seems very fine tuned
from the particle physics point of view and so the
prejudice in a large part of the particle physics community has been
that it should
be zero, by some mechanism not yet understood.
An extremely small but nonzero $\Lambda$
could perhaps be taken as some small effect perturbing
around the preferred value of zero.

Before listing some particle theory mechanisms which have been suggested,
it should be noted that many variants of inflation
(Guth 1981, Linde 1982, Albrecht \& Steinhardt 1982)
correspond to an effective $\Lambda \ne 0$.  (A notable
exception is kinetic inflation,
but this has many problems of its own, see Levin (1995) for discussion
and references.)  Inflation occurs when the scale factor accelerates,
$\ddot{a} >0$.
A period of inflation in the early universe
could have produced the observed homogeneity and
isotropy seen today, as well as seeding density perturbations
for structure formation.  Slow roll inflation, for example,
occurs when a field has approximately constant potential energy.
This approximately constant potential energy behaves like a
cosmological constant, eventually producing inflationary expansion.
So in this inflationary model $\Lambda$ was
effectively nonzero in the past at some time.
If there was an effective nonzero $\Lambda$ in the past, by this or any
other mechanism,
any general prediction giving small or vanishing $\Lambda$
now would have to allow for the earlier exception as well.
Inflation generically drives $\Omega_k \to 0$, or $\Omega_\rho +
\Omega_\Lambda \to 1$
(Peebles 1984, Turner, Steigman \& Krauss 1984).
It does not fix $\Omega_\rho$ or $\Omega_\Lambda$
separately.

There are several arguments, none as yet compelling,
 setting $\Lambda$ to zero or a very small number.  These
include (based on a list by J. Lykken (1998)):
\begin{itemize}
\item Set the cosmological constant to its
current value by fiat.
\item Say that there are many possible universes
with different values of $\Lambda$, but that any measurement
we make has as the prior condition that we exist.  This prior
condition can be restated in terms of galaxies existing or
other objects.  Some references for these anthropic
arguments are given in CPT.  Recent developments include
astrophysical versions producing $\Omega_\Lambda$ of similar size
to $\Omega_\rho$
(for example Efstathiou 1995, Martel, Shapiro \& Weinberg (1998)
and references therein) and refinements of the
probability measure for eternal inflation
(for example Linde, Linde \& Mezhlumian 1995, Vilenkin 1998
and references therein).
\item Wormholes (nucleation of baby universes) might provide a dynamical
potential which forces $\Lambda $ to zero, although the arguments haven't
worked so far (Giddings and Strominger 1988, Coleman 1988, see CPT
for more discussion).
\item Some symmetry that isn't obvious might set $\Lambda$ to zero, for example
a string theoretic symmetry of the partition function
(Moore 1987, Dienes 1990a,b) or a symmetry of some spacetime
related to
our spacetime via stringy symmetries (Witten 1995).
Some modification, perhaps a small breaking of these symmetries, would
be necessary to produce a small $\Lambda \ne 0$.
\item A small value of $\Lambda$ may appear from the ratio
of two numbers of very different size (hierarchy) already present in
particle physics.  An example of a small ratio is
the mass of a very light particle
over the Planck mass, nonperturbative effects in
particle theory also can provide small numbers of
the form $e^{-const./\alpha}$ with $\alpha$ the coupling.
\item The vacuum energy predicted above was found by
counting the degrees of freedom in the context of field theory.
In this case the number of modes
is determined by the volume of space.
This is in contrast to string theory, a candidate for combining
gravity and quantum field theory, where the holographic hypothesis
(Susskind 1994) states that a volume
can be described by properties of its boundary.
The number of degrees of freedom is then limited by the area of
the boundary, not the volume of the space contained within.
In the context of this hypothesis, the corresponding $\rho_{vac}$
has been addressed by Banks (1995).  An estimate of $\rho_{vac}$
via a similar argument
(using limits on black hole entropy to reduce the number of
degrees of freedom) has also been given by Cohen, Kaplan, \& Nelson (1998).
\end{itemize}

\section{kinematics}
Kinematics with nonzero $\Omega_\Lambda$
is summarized in equation \ref{aeom} above for the scale factor.
This section will specialize to nonrelativistic matter,
$\Omega_\rho = \Omega_{NR}$ for simplicity.  The generalizations
from including the other terms in $\Omega_\rho$ are more involved but
immediate by substitution.
With only nonrelativistic matter,
equations (\ref{aeom}, \ref{Ez}) become:
\eq
\frac{da}{dt} = H_0 (\Omega_{NR}a^{-1} + \Omega_k
+ \Omega_\Lambda{a^2})^{1/2} \; .
\label{nreom}
\en
As $a$ is increasing in an expanding universe, the factor
multiplying $\Omega_\Lambda$ increases with time.
If $a$ keeps increasing, $\Omega_\Lambda$ eventually dominates.
Conversely,
at early times, $a \ll 1$, and so the contribution proportional to
$\Omega_\Lambda$ is small.  Even at the relatively recent redshift of
$z=10$ the contribution of $\Omega_\Lambda$ is suppressed by
$10^{-3}$ relative to that of $\Omega_{NR}$.
Note that since
the sum of $\Omega_\rho +  \Omega_\Lambda + \Omega_k = 1$,
changes in $\Omega_\Lambda$ can also be interpreted as changes in
$\Omega_\rho$ and/or $\Omega_k$.

Equation \ref{nreom} can be used to calculate the age of the universe.
The age is the integral of the time up
to now in terms of the scale factor $a$:
\eq
H_0 \int dt = H_0 \int \frac{dt}{da} da =
\int \frac{da} {(\Omega_{NR}a^{-1} + \Omega_k + \Omega_\Lambda a^2)^{1/2}}
\en
Increasing $\Lambda$ increases the age.  This is one reason that
cosmologists favoring $\Omega_k =0$ (i.e. flatness, from inflation)
wanted to introduce a cosmological
constant when high $H_0$ observations were made.  A high
$H_0$ and lower limit on the age of the universe result in a lower
limit on the left hand side of the above equation.  Raising
$\Omega_\Lambda$ raises the value of the right hand side to 
produce agreement.  Other observations, including age measurements,
have also been interpreted as requiring nonzero $\Lambda$ even
without demanding flatness (for example, Gunn \& Tinsley 1975).
In figure one below, one sees that $H_0$ times the age
for a flat ($\Omega_k =0$) universe increases as
$\Omega_{NR} = 1-\Omega_\Lambda$ decreases.
\begin{figure}
\leavevmode\epsfysize=5.4cm \epsfbox{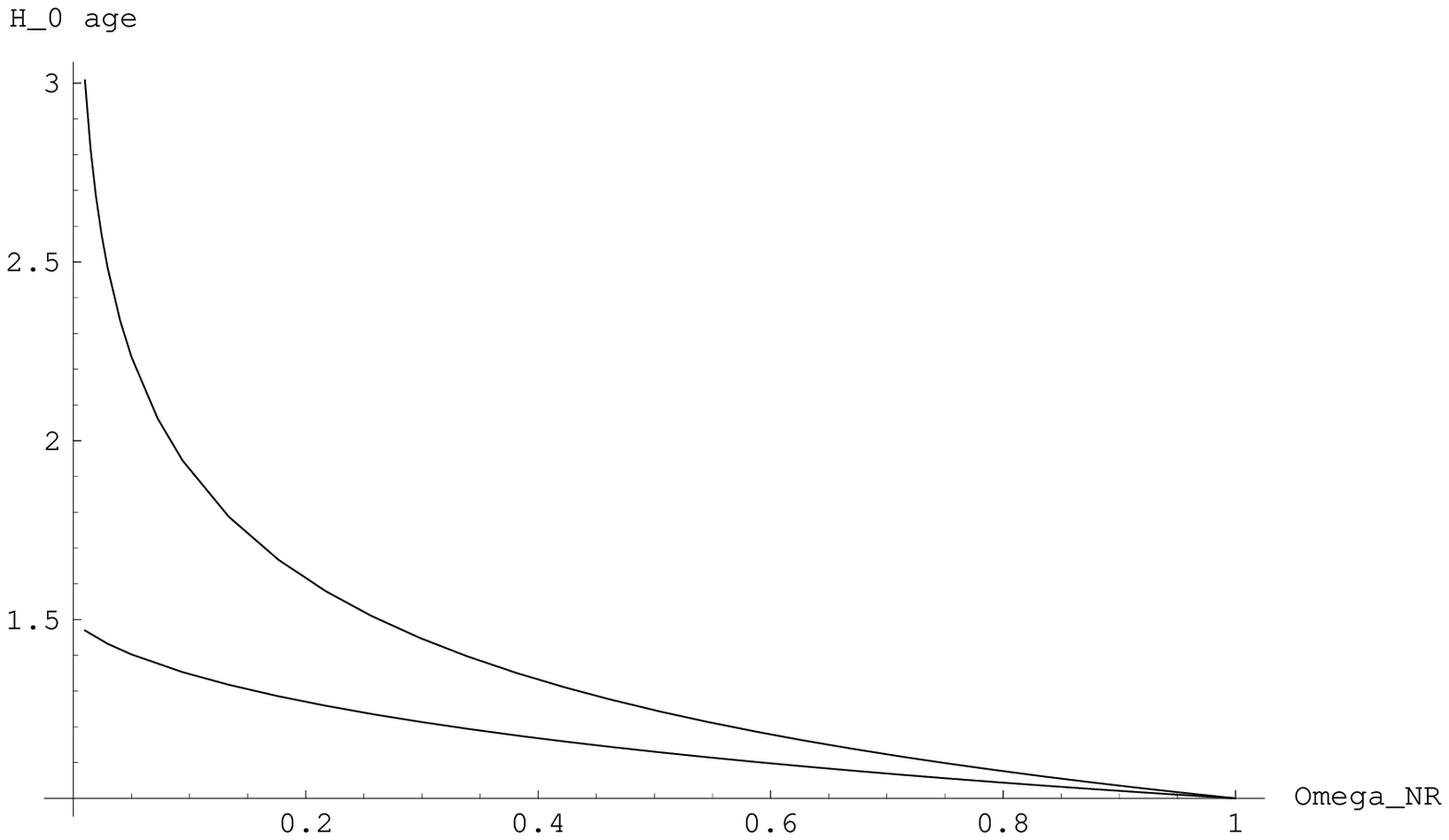}
\leavevmode\epsfysize=5.4cm \epsfbox{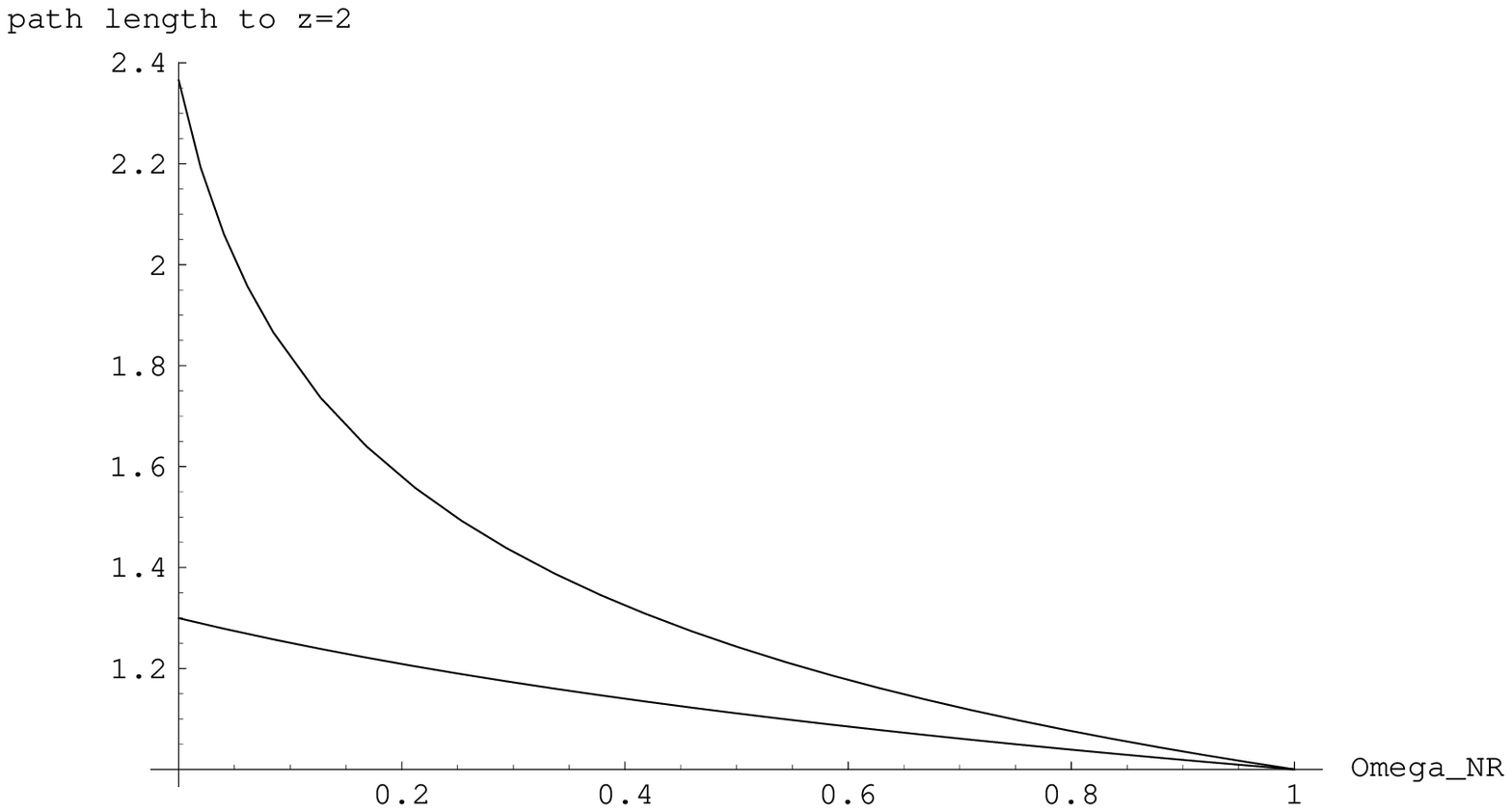}
\vspace{10pt}
{\caption{\small
Age and path length as a function of $\Omega_{NR}$.
 On the left, the top curve shows $H_0$ times the age
for a flat universe, $\Omega_\Lambda + \Omega_{NR}
= 1$, plotted versus $\Omega_{NR}$ and
normalized to one for $\Omega_{NR} = 1$.  The lower curve
shows the same quantities
for an open universe, $\Omega_{NR} + \Omega_k = 1,
\Omega_\Lambda = 0$.  (The effects of
a nonzero but small $\Omega_R$ are negligible in both.)
Shown at right is the path length $\chi$ (equation (20))
of light
back to redshift $z=2$ relative
to the path length for an $\Omega_{NR}=1$ universe, again
plotted as a function of $\Omega_{NR}$.  The upper curve
is for
a flat universe, $\Omega_{NR} + \Omega_\Lambda =1$,
and  the lower curve is for
an open universe, $\Omega_{NR} + \Omega_k = 1$.}}
\end{figure}

The distance a light ray travels can be calculated as well.
Light rays follow null geodesics where
$ds^2 =0$ so that $dt^2 = R_0^2 a^2 d\chi^2$.
Then
\eq
\int R_0 d \chi
= \int \frac{dt}{a(t)} = \int  da \, \frac{1}{a}  \frac{dt}{da}
\en
and so the path back to scale factor $a$ or redshift $z$ is
\eq
\chi(a) =
 \int_a^1 \frac{da'}{R_0\dot{a'} a'}
=  \int_0^{z = a^{-1} -1} \frac{dz'}{R_0 H_0 E(z')} \; .
\label{ra}
\en
This path length $\chi$, for light going back to
redshift $z=2$, is shown above for a flat $\Omega_k = 0$ universe and
for an open universe with $\Omega_\Lambda =0$,
as a function of varying $\Omega_{NR}$.  The path length is
relevant for the measurements of lensing mentioned later.
Another astrophysical quantity is
the angular diameter distance $d_A$,
the ratio of the proper size of an object to its apparent angular size.
In this notation,
\eq
d_A = \frac{R_0 r(\chi)}{1+z} = \frac{1}{H_0 \sqrt{|\Omega_k|}}
\frac{r(\chi)}{1+z}
\label{angd}
\en
and the luminosity distance, related to the known rest frame luminosity
and the apparent flux, is
$d_L = (1+z)^2 d_A$. (Recall that $c=1$ in these units.)

One can also calculate the volume  seen looking back to a
given redshift.  The comoving volume is
\eq
dV = R_0^3 r^2 (\chi) d\chi d \Omega
\label{volume}
\en
and so using the equations above for $r(\chi)$ and
$d\chi$ this can be calculated directly.  The volume effects
are relevant for measurements of number counts of objects such as
gravitational lenses.
More plots of the kinematic effects of changing $\Omega_\Lambda$
are found in CPT.

\section{Structure Formation}
Structure formation proceeds by
gravitational collapse of small amplitude density
perturbations into larger and
larger ones.  A cosmological constant
corresponds to an energy density which is smooth, that is,
doesn't clump or change on any scale.  Thus, for a
given $\Omega_k$, increasing $\Omega_\Lambda$ decreases $\Omega_\rho$,
the matter available for gravitational collapse.
Consequently structure grows more slowly in a universe with
larger $\Omega_\Lambda$ for fixed $\Omega_k$.  As inflationary models
can most easily provide $\Omega_k=0$, comparisons often fix
$\Omega_\Lambda + \Omega_\rho = 1$ to get intuition into the effects of
$\Omega_\Lambda$.

One can be slightly more precise by looking at the growth of
density perturbations.  The homogeneity of, 
for example, the CMB, tells us that
at early times fluctuations had very small amplitude,
of order parts in $10^{-5}$.  In this case linear perturbation theory
is expected to work very well.  At later times they can be followed
into the nonlinear regime with semi-analytic methods or numerical
simulations.  Fluctuations on the largest scales have the smallest amplitude
because they have only recently entered the horizon and thus only recently
started to collapse.  Hence on large scales, for example for galaxy
redshift surveys, linear theory is useful
for direct study of fluctuations.

For smaller scales, fluctuations have a larger amplitude and
hence nonlinearities begin to be important.  The scale for this
crossover today is somewhere around 10 $h^{-1}$ Mpc.
In this case one can instead consider number densities.
A good fit to the results of numerical simulations is
to predict the number of regions with $\delta > \delta_c \sim 1.7$ and then
model the collapse of these regions assuming spherical symmetry.
This Press-Schechter (1974) method
actually works well outside the expected range of validity;
the excellent agreement of its results with simulations
motivates its use.
It can be considered as a very good analytic fit to simulation results.
One application which will be discussed later is
the rich cluster abundance.

The effects of a cosmological constant on linear perturbation theory
is as follows.
Over short distances, where the
Newtonian limit is applicable,
the equation of motion for density fluctuations of type $A$,
$\delta \rho_A/\rho_A$ is (for example Padmanhabhan 1993)
\eq
\ddot{\delta}_A + 2 \frac{\dot{a}}{a} \dot{\delta}_A -
\frac{v_A^2}{a^2} k^2 \delta_A  =
4 \pi G \sum_B \rho_B \delta_B
\label{flucts}
\en
where $v_A$ is the sound speed.
The sum on the right is over all components $\delta_B$,
but for $\rho_\Lambda$,
$\delta_\Lambda=0$ because the cosmological constant doesn't have
any associated fluctuations.  Nonzero $\Omega_k$, curvature,
also makes no contribution to the right hand side.

A simple case is when only one type of nonrelativistic
energy density $\rho_C$ dominates the fluctuations,
$v_C \propto p_C/\rho_C = 0$.
For redshifts where this
$\Omega_{NR}$ dominates, equation (\ref{flucts}) becomes
\eq
\ddot{\delta}_C + 2\frac{\dot{a}}{a}\dot{\delta}_C =
4 \pi G\rho_C \delta_C  \; \; ({\rm matter \; domination}) \; .
\label{nrflucts}
\en
First
ignore the expansion of the universe (the $\dot{a}/a$ term) and
the time dependence in $\rho_C$.
Then the growing solution is
$ \delta_C \sim e^{\sqrt{4 \pi G \rho_C} \, t} $, illustrating that
increasing $\rho_C$ increases the rate of fluctuation growth.
This is to be expected since the right hand side is
the change in energy density,
the gradient of the gravitational potential.
Including the expansion of the universe, the
$\frac{\dot{a}}{a} \dot{\delta}$ term
acts as a drag term to slow the growth of the perturbations,
from exponential growth to power law, but the trend of
faster growth with larger $\rho_C$ remains.
On the other hand,
with only one sort of matter as assumed here,
increasing $\Omega_\Lambda$ for fixed curvature $\Omega_k$
means decreasing $\rho_C$.
Thus here increasing $\Omega_\Lambda$ means perturbations
grow more slowly.

In the limit that the source term $\rho_C \delta_C$ is very small,
the expansion of the
universe prevents structure from growing.  The limit
$\rho_C = 0$ gives
$\delta_C \sim constant$.
The drag coefficient from the expansion of the universe is
given by (equation (\ref{nreom}) above)
\eq
\frac{\dot{a}}{a}=
H_0 (\frac{\Omega_{NR}}{a^3} + \frac{\Omega_k}{a^2} +
\Omega_\Lambda)^{1/2} \; ,
\en
while in the source term
the nonrelativistic density $\rho_C$ drops as $a^{-3}$.
So if either $\Omega_k$ or $\Omega_\Lambda$ is nonzero,
with increasing scale factor $a$, the drag term
$\frac{\dot{a}}{a} \dot{\delta}_C$ will become larger than
the source term $4 \pi G \rho_C \delta_C$, and
structure will eventually stop growing.
The extreme case of this
is when the cosmological constant dominates at late times,
making the universe inflate.
Then,
regions which have broken away via gravitational collapse will
continue to evolve separately (hence the term breaking away),
but these regions will be separated from each other more and more
as time goes on.

Equation (\ref{nrflucts}) has a formal solution.  In this
equation, make the substituion
\eq
\frac{d}{dt} = \dot{a} \frac{d}{da} = a H \frac{d}{da} \; ,
\en
define $\delta^\prime = d \delta/da$, etc., and
define
$\delta(a)$ is implicitly by $\delta(t)$ and vice
versa via $a(t)$.
Dropping the subscript $C$,
the fluctuation equation (\ref{nrflucts}) then becomes
\eq
a^2 H^2 \delta '' +
a (3 H^2 + a H H') \delta ' = 4 \pi G \frac{\rho_{NR}}{a^3} \delta \; ,
\en
with solution (Heath 1977)
\eq
\delta (a) = \frac{5}{2} H_0^2 \Omega_{NR} \frac{\dot{a}}{a} \int_0^a
\frac{d \tilde{a}}{\dot{\tilde {a}}^3}  \; .
\en
This has been normalized to give
$\delta(a) \sim a$ in the case $\Omega_{NR}=1,
\Omega_\Lambda = 0$.
  The behavior of the solution $\delta (a)$
is shown in Figure 2
below for two cases, $\Omega_{NR} =1, \Omega_\Lambda =0$ and
$\Omega_{NR} = .3, \Omega_\Lambda = .7$.
The relative size of the fluctuations in different cosmologies today is
given by $\delta(a=1)$.
\begin{figure}
\centerline{\epsfig{file=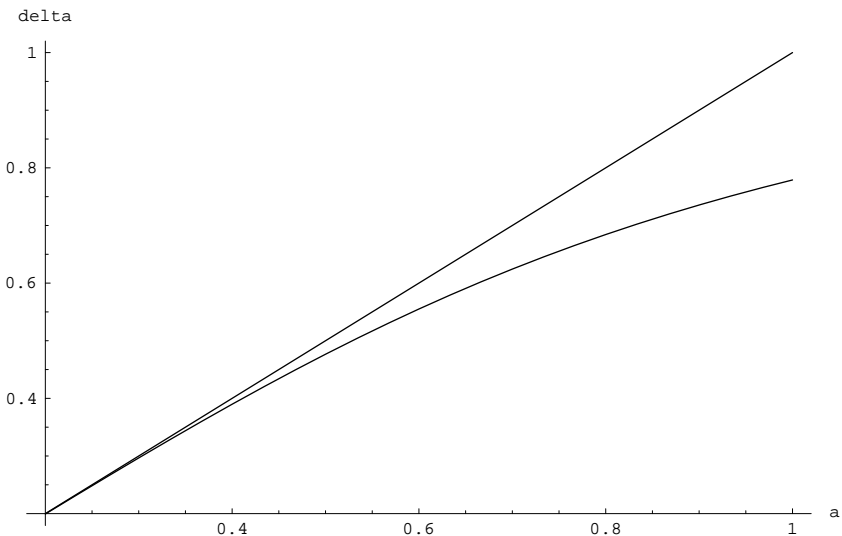,height=3.0in}}
\vspace{10pt}
{\caption{\small
Growth of perturbations as a function of $a$
for an $\Omega_{NR}=1, \Omega_\Lambda =0$
model (where $\delta(a) \sim a$) and for a
$\Omega_{NR} = .3, \Omega_\Lambda = .7$ model (lower curve).}}
\end{figure}

Thus, in summary,
both curvature and the cosmological constant have no fluctuations,
corresponding to smooth components,
$\delta_{curv} = \delta_\Lambda =0$.  Consequently
fluctuations grow more slowly when there is a positive
cosmological constant, or if there is curvature $\Omega_k > 0$.

Peculiar velocities also respond to gravitational potentials and
hence the growth of perturbations, but the rate of growth and
peculiar velocities today ($z \sim 0$) are only very weakly dependent
upon the cosmological constant, as are many other
dynamical measurements (Lahav et al 1991, CPT).  For example, the
growth of perturbations at the present epoch goes as
$\sim \Omega_{NR}^{0.6} +
\frac{1}{70} \Omega_\Lambda(1 + \Omega_{NR})$.

\section{Measurements}
Recently there have been significant
observational advances relevant to measuring the cosmological
constant.  These include the following.

{\bf Gravitational Lenses}:

Gravitational lensing of quasars by intervening galaxies
measures the volume of space back to a given redshift, assuming
constant comoving density of lensing objects.
The volume back to a given redshift (derived from
equation (\ref{volume})) has a strong dependence
on the cosmological constant
in a flat universe (Fukugita et al 1990, Turner 1990,
much of the analysis was implicit in Gott et al 1989).
The lens density relative
to the fiducial case $\Omega_{NR} = 1, \Omega_\Lambda =0$ is
(Fukugita et al 1992, CPT):
\eq
P_{lens} = \frac{15}{4} (1 - \frac{1}{\sqrt{1 + z_s}})^{-3}
\int_0^{z_s} dz
\frac{ (1+z)^2}{E(z)} H_0^2
\left[\frac{d_A(0,z)d_A(z,z_s)}{d_A(0,z_s)} \right]^2
\en
for a source at redshift $z_s$.  Recall that $c=1$ so that $H_0 R_0$
is dimensionless.  The
angular distance $d_A(z_1,z_2)$ between redshifts $z_1,z_2$ is
the generalization of equation (\ref{angd}):
\eq
d(z_1,z_2) = \frac{R_0 }{(1 + z_2)}
{\rm sinn} \{ \int_{z_1}^{z_2} dz \frac{1}{R_0 H_0 E(z)} \}
\en

In Figure 3, $P_{lens}$
is plotted for a flat ($\Omega_k = 0$)
universe for sources at $z_s =2$.
The effect of a large $\Lambda$ in a flat universe is very
strong for $\Omega_{NR} < .3$, i.e. $\Omega_{\Lambda} > .7$.
The results of integrating the above for
a range of different $\Omega_k$ are found in CPT.
\begin{figure}
\centerline{\epsfig{file=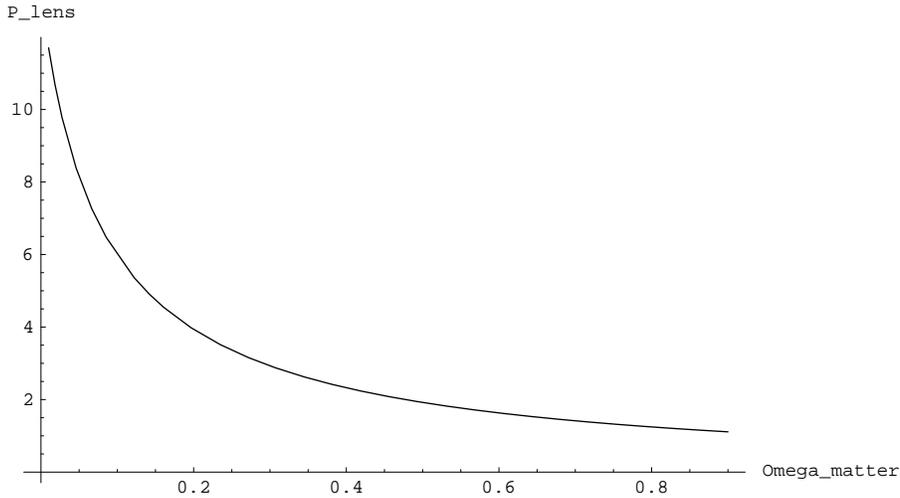,height=3.0in}}
\vspace{10pt}
{\caption{\small
Number of lenses expected for sources at z=2
in a flat universe with varying $\Omega_{matter}=
\Omega_{NR}$, relative to
the number expected for $\Omega_{matter}=1$.}}
\end{figure}

The current situation with the data is as follows.
For $\Omega_k=0$, analysis (Kochanek 1993, 1996; Maoz \& Rix 1993)
of surveys for
multiply imaged quasars give a 2$\sigma$ upper limit of
$\Omega_\Lambda < .66$.  There are statistical errors because
of the small number of lensed quasars in the survey and
uncertainties in the local number counts of
galaxies by type.  Possible sources of
systematic errors (for a full list of references see
Falco, Kochanek \& Munoz 1998)
include extinction, galaxy evolution, the quasar discovery
process and the model for the lens galaxies.  Using radio selected lenses
to reduce possible systematic errors associated with
extinction and the quasar discovery process, Falco et al
find $\Omega_\Lambda < .73$ at 2$\sigma$.   Combining the radio and optical
data they find $\Omega_\Lambda < .62$ at $2 \sigma$ for their
most conservative model.
The errors should improve significantly with the upcoming
observations, including for example the
CASTLe survey (the CfA-Arizona Space
Telescope Lens Survey)\footnote{http://cfa-www.harvard.edu/castles},
which will have an HST survey to measure redshifts of lens galaxies,
and the completion of the CLASS (Cosmic Lens All-Sky Survey)\footnote{
http://dept.physics.upenn.edu/~myers/class.html}
 radio lens survey.

\bigskip

{\bf Cluster abundance}:

In the previous section it was argued that fluctuations grow more quickly
for larger $\Omega_\rho$.
Thus, fixing $\Omega_k$ and
increasing $\Omega_\Lambda$, perturbations will grow more and more slowly.
Galaxy clusters are rare objects on large scales which have only
recently gone nonlinear, and thus many of the complications due to
nonlinearities may be expected to be less important.
Consequently, normalizing to the rich cluster
abundance seen today at 8 $h^{-1}$ Mpc, $\sigma_8$, and fixing
$\Omega_k$,
a higher $\Omega_\Lambda$ universe has earlier cluster
formation (i.e. at higher redshift).

An example of the different number of clusters expected as a function
of cosmology is shown below.
\begin{figure}
\centerline{\epsfig{file=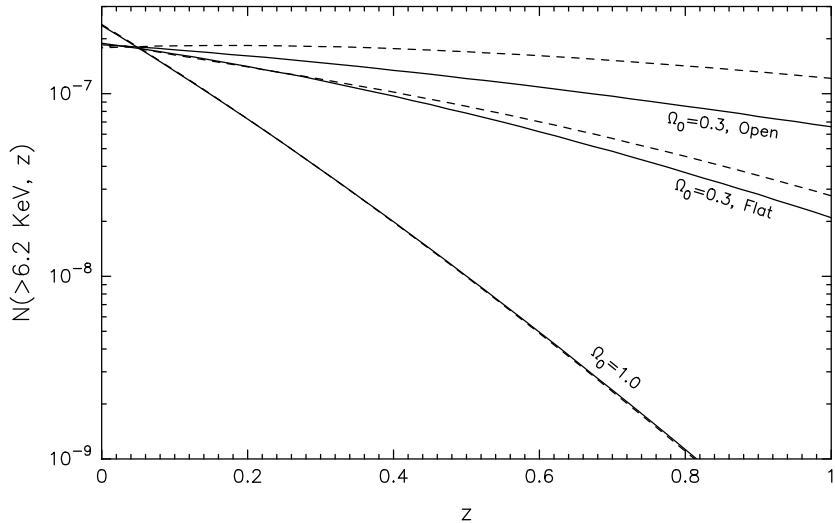,height=3.0in}}
\vspace{10pt}
{\caption{\small
Expected redshift evolution of
$N(> 6.2 {\rm keV},z)$, the
comoving number density of galaxy clusters with an X-ray
temperature $k_B T >  6.2$ keV at redshift $z$, normalized to
produced the observed  $N(>6.2 {\rm keV},0.05)$.  The cluster virialization
redshift $z_c$ was estimated
using a method of Lacey \&
Cole (1993, 1994) for the
solid lines, and $z_c$
was taken to coincide with the cluster redshift for the dashed lines.
The $\Omega_0=0.3$ flat model has  $\Omega_\Lambda = 0.7$.
Courtesy of Viana and Liddle.}}
\end{figure}

To compare observations with theory,
Press-Schechter or N-body calculations
are used to calculate the expected mass distribution,
and then the mass is related to the X-ray temperature.
The relation between the mass and the X-ray temperature is one
of the major uncertainties.
The numerically calculated cluster number evolution can then
be compared to current data.  A wide range of results are found in
Eke et al (1998), Viana \& Liddle (1998), Reichart et al (1998),
Blanchard \& Bartlett (1997),
Henry (1997), Fan, Bahcall \& Cen (1997), Gross et al (1997), Henry (1997)
and references therein.

The deepest complete X-ray sample available is from the
Einstein Medium Sensitivity Survey (EMSS)\footnote{
http://www.tac.dk/~lars\_c/gammabox/doc/emss.html}.
The statistics and uncertainties
(both in modeling and in the data)
are not yet well enough under control to determine $\Omega_\rho$
conclusively (Colafrancesco, Mazzotta \& Vittorio 1997,
Viana \& Liddle 1998), although
the mere presence of
very high ($z >1$) redshift clusters has been argued sufficient to rule out
$\Omega_\rho = 1$ (for example, Gioia 1997, Bahcall \& Fan 1998b
and references therein).

Several new surveys are underway in the X-ray and optical,
and
upcoming satellites such as AXAF (the Advanced X-ray
Astrophysics Facility)\footnote{
http://asc.harvard.edu/} and
XMM (X-ray Multi-mirror
Mission)\footnote{http://astro.estec.esa.nl/XMM/xmm.html}
 should allow better determinations
of cluster properties.  Thus observational data should markedly improve
in the very near future.

\bigskip

{\bf Arcs}:

The number of galaxy clusters at a given time also affects the number of
strong gravitational lenses (i.e. arcs).
Recently Bartelmann et al (1997,1998, see references therein
for earlier work) analyzed the
number of lenses expected in different cosmologies.
For background sources at $z_s \sim 1$,
clusters at redshifts $0.2 \le z_c \le 0.4$ are
the most efficient lenses, with only weak dependence on cosmology.
More lenses are expected for lower $\Omega_\rho$ (and hence
higher $\Omega_k$ and $\Omega_\Lambda$) because
in this case the clusters form earlier and thus are
more likely to be present to act as lenses.  In addition,
they argue that clusters forming at an earlier time are expected to be
composed of more compact subclusters and thus be more
efficient at lensing.

Using two cluster simulation methods and varying the cosmological
parameters (nine simulations in all), Bartelmann et al (1997, 1998)
found an order of magnitude more arcs are produced in a
flat $\Omega_\Lambda = 0.7$ than in a flat $\Omega_\Lambda =0$
universe.  (For an open $\Omega_\rho = 0.3, \; \Omega_\Lambda =0$
universe they expect an additional factor of ten more.)
The number of arcs estimated from extrapolating the
EMSS (Einstein X-ray Extended Medium
Sensitivity Survey) arc survey to the
whole sky is about 1500-2300, while the number of arcs expected from
their simulations is
\eq
N_{arcs} \sim \left\{
\begin{array}{lll}
2400 \; \; \; &\Omega_\rho = .3 \; \; \; & \Omega_\Lambda = 0 \\
280 & \Omega_\rho = .3 & \Omega_\Lambda = .7 \\
36 & \Omega_\rho = 1 & \Omega_\Lambda = 0
\end{array}
\right.
\en
As with most observations described here, the data is
expected to improve significantly.

\bigskip

{\bf Radial/Transverse distance ratios}:

A method which shows much promise for investigating $\Omega_\Lambda$
in the future is related to the
distortion of quantities which are intrinsically isotropic
(Alcock \& Paczynski 1979).
Transverse and radial distances have a different dependence
on $\Omega_\Lambda$, if they are equal this is a condition
on cosmological parameters.  In more detail,
{}from the metric, equation (\ref{metric}), the comoving distance
between two objects is
\eq
d\chi^2 + r^2(\chi) [d \theta^2 +
\sin^2 \theta d \phi^2] = \left(\frac{d \chi}{dz} \right)^2
dz^2 + r^2(\chi)[d \theta^2 +
\sin^2 \theta d \phi^2] \; ,
\en
where $d \chi $ has been rewritten as $\frac {d \chi}{d z} dz$.
With an isotropic spacetime, something which is intrinsically
isotropic should be measured to have the same size in all directions.
For example, the measured distance in the redshift (radial) direction
should be the same as
the measured distance in the angular ($\theta$) direction:
\eq
\frac {d \chi}{d z} dz = r(\chi) d \theta \; .
\en
As $d\chi/d z = 1/(R_0 H_0 E(z))$ from equation (\ref{ra}),
and $r(\chi)$ is given by equations (\ref{rchi}, \ref{ra}),
one expects for an intrinsically isotropic object that
\eq
r(\chi) R_0 H_0 E(z) = \frac{dz}{d \theta} \; .
\en
The left hand side is calculable for a given
model while the right hand side is measured.
The left hand side can also be written in terms of the
angular diameter distance $d_A$ given earlier.  The
notation $f/g$ is used in Phillipps (1994) and subsequent papers.

Although galaxy clusters, the original candidate for isotropic
objects,
are not intrinsically isotropic enough for this test, the correlation
function of, for instance, quasar pairs (Phillips 1994), is
expected to be intrinsically isotropic
when averaged over pairs.
The feasibility of using this particular test with the Sloan Digital Sky
Survey\footnote{http://www.astro.princeton.edu/BOOK} and
the Two Degree Field Survey\footnote{http://meteor.anu.edu.au/~colless/2dF}
has been analyzed by
Popowski et al (1998), other suggested uses of this test are
also referenced and discussed therein.

\bigskip

{\bf Supernovae}:

Type Ia supernovae are considered to be calibrated candles (once
a width-brightness relation is applied, their absolute magnitude $M$ is
believed to be known).  Their apparent magnitude $m$ is given by
\eq
m-M = 5 \log_{10} \frac{d_L}{\rm Mpc} + 25
\en
where $d_L$, the luminosity distance, is given by
\eq
d_L =R_0 (1+z) {\rm sinn} \; \int_0^z \frac{dz'}{R_0 H_0 E(z')} \; .
\en
Thus the curve for  $d_L(z)$ versus $z$ depends upon cosmological
parameters.

Using a method pioneered by Perlmutter et al (1995), two groups
(Perlmutter et al 1998, Riess et al 1998 and references therein)
have been steadily finding and analyzing supernovae for the
past couple of years.  The data is accumulating rapidly.
As of this
writing, $d_L(z)$ versus $z$
results for several dozen high redshift supernovae have been
announced,
and light curves of many others are being followed, several of which
will have HST observations.
The combined results have so far indicated high confidence
for a positive nonzero cosmological constant.  For the most recent
information their web pages can be consulted\footnote{
http://www-supernova.lbl.gov/ and
http://cfa-www.harvard.edu/cfa/oir/Research/supernova/HighZ.html}.

Possible systematic errors which have been studied include
extinction and evolution, selection effects, weak lensing,
the width-brightness relation used for calibration and sample
contamination.  In particular,
there is very little
reddening seen in either the far or near supernovae.  There are models
which predict this, for example the Monte Carlo analysis by
Hatano et al (1997) on observability of supernovae 
predicts ``the probability distribution of
extinction for type Ia supernovae to be strongly peaked near zero.''
As so much relies upon SN Ia being standard candles,
the question of evolution is crucial as well.
Spectra of the high and low redshift supernovae have been compared with
good agreement.  The distribution of faint and bright
supernovae appears similar at both high and low redshift.
Many of the uncertainties now come from the low redshift samples.
It is expected that there will be improvements in understanding the
calibrations as the number of supernovae Ia found in
galaxies where there are
Cepheids increases (Kirshner 1998).

\bigskip

{\bf CMB anisotropies}:

The study of
CMB anisotropies provides a useful complement to the
supernovae results.
The supernovae results at redshift $z \sim .5$
measure $\Omega_\rho - \Omega_\Lambda$.  In contrast,
the CMB anisotropy measurements yield a power spectrum
as a function
of angle.  Thus, positions of features in the spectrum,
in particular the first doppler peak
(White and Scott 1996), measure the angular diameter distance
(equation (\ref{angd}), which is a function of the weighted sum of
$\Omega_\rho + \Omega_\Lambda$.
Combining the results from both types of studies is
a promising tool
(White 1998, Tegmark, Eisenstein \& Hu 1998,
Tegmark et al 1998b,
Garnavich et al 1998).  The current best fit to the existing data
is (Lineweaver 1998)
\eq
\Omega_\Lambda = .62 \pm .16
\; , \; \; \Omega_\rho = .24 \pm .10
\en
where the errors denote 68.3\% confidence level.
A likelihood plot of supernovae and CMB data is shown in figure
five below.  The current situation is on the left, while
the expected
precision for a particular high $\Omega_\Lambda$
model after the flight of the Planck
surveyor\footnote{http://astro.estec.esa.nl/Planck/.  Planck will
also be a source of relevant data for other tests, for example lenses
(Blain 1998 and references therein).} is shown at right.
\begin{figure}
\center{
\leavevmode\epsfysize=5.4cm \epsfbox{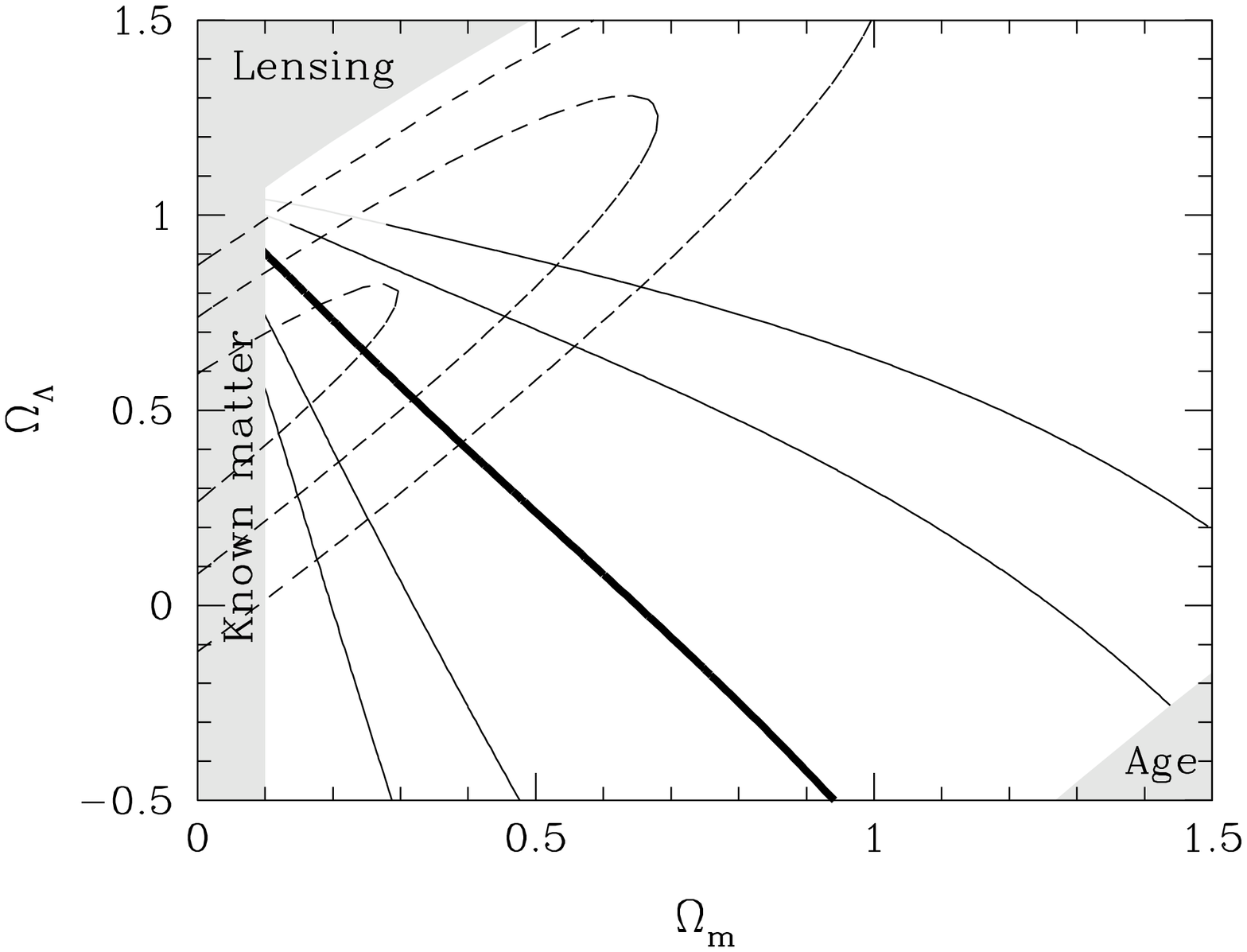}
\leavevmode\epsfysize=5.4cm \epsfbox{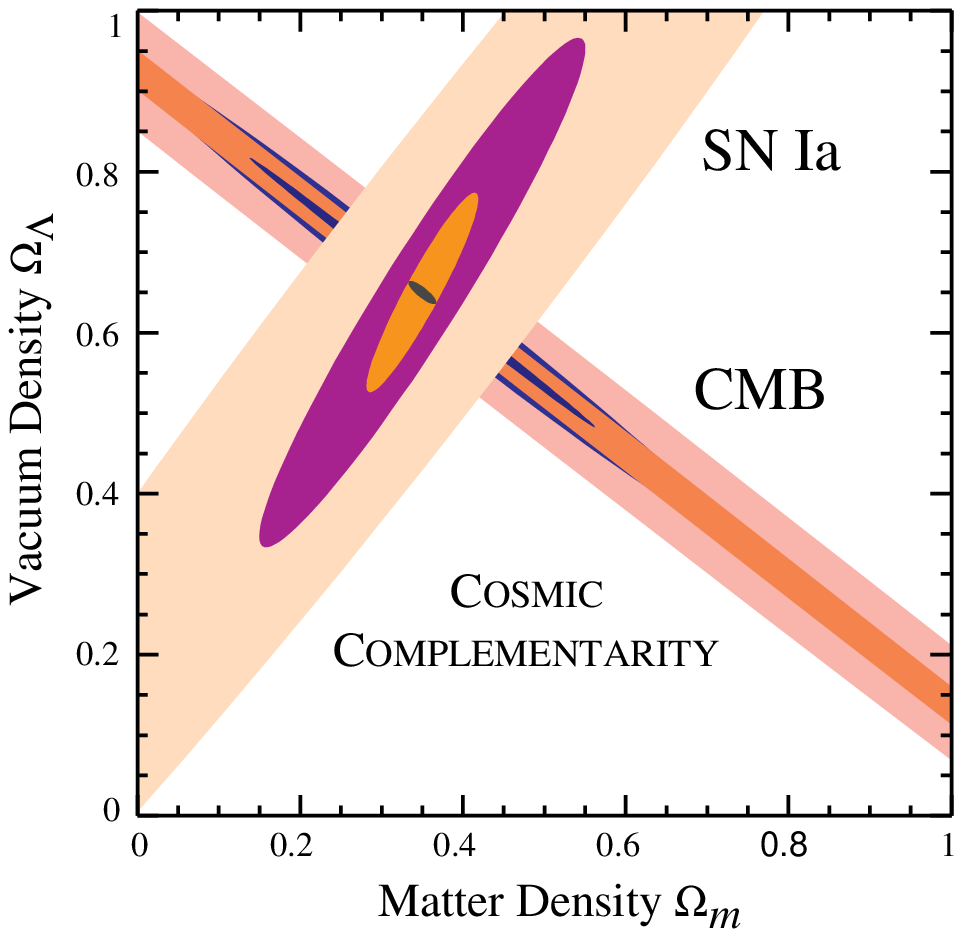} }
\vspace{10pt}
{\caption{\small
Cosmic complementarity:
likelihood in the $\Omega_m$-$\Omega_\Lambda$ plane
for fitting both SN-Ia and CMB data.  On the left is
compilation of the current data, courtesy M. White.
The dashed lines are $1\sigma$, $2\sigma$ and $3\sigma$
contours for fitting
the SN-Ia results, the solid lines are the CMB results.  The thick
solid contour denotes the peak of the likelihood found by
Hancock et al (1998), and the two contours to either
side represent conservative $\pm1\sigma$ and $\pm2\sigma$
values.
The shaded areas are ruled out by other constraints as indicated.
On the right is shown, courtesy Eisenstein \& Hu,
the precision expected with
CMB data from upcoming Planck satellite mission
and `optimistic' SN-Ia data, for a model with $\Omega_{NR} = 0.35$,
$\Omega_\Lambda = 0.65$.  One $\sigma$ confidence regions are shown and the
combined data has an error region the size of the overlap.  The
central bar gives the errors expected with polarization included.
For more discussion see Tegmark et al (1998b).
}}
\end{figure}

As the CMB anisotropy data will be improving
in upcoming years, so will these tests.
The CMB
can also be used in conjunction with measures of $\Omega_\rho$,
such as mass/light ratios (see this recent review by
Bahcall \& Fan (1998a)) or peculiar velocities of
galaxy clusters.  For example,
low $\Omega_\rho$ combined with CMB evidence
for a flat universe would support $\Omega_\Lambda \ne 0$.

\bigskip

{\bf Other measures}:

There are several other tests,
including
number counts, QSO line statistics, galaxy mergers,
weak lensing surveys (for example Van Waerbeke, Bernardeau \& Mellier 1998
and references therein),
etc., CPT discuss several of these.  Some tests, such as
number counts, are now understood to be less reliable than earlier believed
due to evolutionary effects.

\section{Maybe not $\Lambda$}
Although a cosmological constant has been suggested as providing
the missing
energy density in the universe, this is not the only possible
explanation of the observations.
As mentioned earlier,
the extremely small but
nonzero cosmological constant
$\Lambda$ consistent with current data is
theoretically unnatural given the current understanding of particle
physics.  This
gives rise to the question of
whether the observations can be explained by
some other form of energy density as well.
In some scenarios this other energy density could replace nonzero
$\Lambda$ entirely, so that $\Omega_\Lambda =0$.
Other forms of energy density have been studied in the past,
in particular by those
wishing to combine the simplest inflationary prediction $\Omega_k \to 0$
with early
observations which favored small $\Omega_{NR}$ and $\Omega_\Lambda$.
Other motivations include the observations that
the galaxy power spectrum amplitude is too high
on all scales for $\Lambda$ models
(for example White \& Scott 1996, Coble et al 1996).

Thus one can postulate some additional energy density scaling as
\eq
\Omega_X \sim a^{-3(1+w)}
\label{xscale}
\en
with $w = p_X/\rho_X$.
For simplicity (again generalization is straightforward),
take only one type of
$\Omega_X$ in addition to the energy densities already considered.
The limit $w \to -1$ gives the cosmological constant.
This additional energy density is called by several
names: quintessence, generalized dark matter, or,
if smooth, X-matter.  It is often specified by a field
theoretic model, and includes models with
a `time dependent cosmological constant'.  
Many candidates for quintessence
have been proposed over the years;
several of the papers previous to 1992 are described and summarized
in CPT,
some more recent descriptions include Coble et al (1996),
Ferreira \& Joyce (1997a, 1997b),
Anderson \& Carroll (1997), Carroll (1998), Zlatev, Wang \&
Steinhardt (1998).  See these papers for additional references
and discussion.
The matter in these field theories sometimes has special properties.
For example, when a scalar field $\phi$
in a shallow potential is used to provide the extra matter,
the field excitations in the simplest case are nearly massless,
$m_\phi \sim 10^{-33} {\rm eV}$.
(This should be compared to limits on the photon's rest mass, which are
in the range $10^{-18} - 10^{-27}$ eV (Nieto 1993).)
Of course,
just as for the cosmological constant, this small number could
be attributed to the ratio of two numbers with very different sizes
already present in particle physics,
for example a scalar depending on the tiny ratio of
neutrino mass/Planck mass (Hill \& Ross 1988).  And more generally,
there are models going beyond the simplest case.

The equation of
state specified by $w$ can be constant or time varying and is related to
the potential in the case of models based on a scalar field.
For tangled strings (Vilenkin 1984, Spergel and Pen 1997) and textures
(Davis 1987, Kamionkowski \& Toumbas 1996) $w=-1/3$.

Including this $\Omega_X$ in Einstein's equations,
equation \ref{aeom} above becomes
\eq
\frac{da}{dt}= H_0(\frac{\Omega_{NR}}{a} +
\frac{\Omega_{R}}{a^2} + \Omega_k +
\Omega_\Lambda a^2 + \frac{ \Omega_X} {a^{1+ 3w}})^{1/2}
\en
The kinematic analysis from
section three can be repeated immediately.  In order to meet
current age and Hubble constant measurements
(Turner \& White 1997, Chiba et al 1998a, Caldwell et al 1998a)
$w < 0$ is preferred.  The general
evolution of $a(t)$ for different equations of state
is given in a recent paper by Overduin and Cooperstock (1998).

To consider effects on structure formation, the behavior of fluctuations
must be specified in addition to the background equation of state.
For example,
a tangled network of light strings will be very rigid, and hence
approximatedly smooth on subhorizon scales.
If $\Omega_X$ is smooth on subhorizon scales, there is less matter
available to fall into density perturbations, and so
in this case structure grows
more slowly.  More generally, a three parameter
stress energy tensor for this matter has
been given in Hu (1998), for the case when the stress fluctuations are
derived from density and velocity perturbations.

Cosmological tests such as CMB predictions, cluster abundance,
supernovae results etc. have been analyzed
for a variety of forms of X-matter.  Some recent papers include
Turner and White (1997),
Caldwell et al (1998a, 1998b), White (1998), Wang \&
Steinhardt (1998), Huey et al(1998), Chiba et al (1998a, 1998b),
Hu et al (1998), Garnavich et al (1998) and references therein.
As there are many parameters which can be tuned in these
generalized dark matter theories, it is possible to find variants which
currently fit observations just as well as the case where
cosmological constant is actually constant.
Just as a search a theoretical explanation of the size of the
cosmological constant
is underway, compelling theoretical explanations for the
presence of generalized dark matter
candidates are also lacking.
Currently it is a question of aesthetics as to which type of matter,
the cosmological constant or generalized dark matter, is a more
reasonable way to provide the
apparently missing energy density.

\section{Summary}
Before summarizing it is worthwhile to remind people of the future of a
universe with large positive cosmological constant.  In this case,
the universe will expand faster and faster, and eventually enter
a stage of inflation, approaching exponential expansion
$a \sim e^{\sqrt{\Lambda/3} \, t}$.  Number densities will drop
and the universe will become more and more homogeneous as time
goes on.  If there is no mechanism to change $\Lambda$, that is,
if the cosmological constant is truly constant, then this
expansion will continue indefinitely.  If, instead, the cosmological
`constant' decreases at some point, then reheating, the filling of
the universe with particles (field fluctuations) may take
place, just as happened in our past if inflation occurred.

To summarize,
there is now strong evidence for accelerated expansion of
the universe from supernovae results.
The supernovae and other measurements are improving rapidly.  Some evidence
which in the past was used to support
$\Omega<1$  is $\Omega_{NR} <1$
evidence, and hence consistent with this picture.
Besides or in place of a cosmological
constant, this observational evidence might also be consistent
with matter having some unusual equation of state.
As such an equation of state
has several parameters, many of the tests (some described above) narrow
the parameter space rather than ruling out this extra form of matter.
Rapid improvement in the quality of the data is expected over the
next few years.

\bigskip

\section{acknowledgements}
I thank C. Kochanek, and S. Perlmutter for explanations about their
work, M. White for discussions, and W. Hu, S. Lamb, A. Liddle,
D. Scott, L. Thompson and M. White
for very useful suggestions on the draft.
I am grateful to S. Lamb for suggesting that this informal talk be
turned into a short pedagogical review, and to the Institute for
Advanced Study for hospitality while this work was completed.
This work was supported by an NSF Career Advancement Award,
NSF-PHY-9722787.
\bigskip

{\bf References}

Albrecht A. \& Steinhardt P.J. 1982, Phys. Rev. Lett. 48, 1220

Alcock C. \& Paczynski B., 1979, Nature 281, 358

Anderson G.W. \& Carroll S.M., 1997, preprint astro-ph/9711288

Bahcall, N. A. \& Fan X., 1998a, preprint to appear
in National Academy of Sciences Proc. 1998, astro-ph 9804082

Bahcall N.A. \& Fan X., 1998b, preprint astro-ph/9803277

Banks T., hep-th/9601151

Bartelmann M., Huss A., Colberg J.M., Jenkins  A.,  Pearce F.R.,
1998, A\& A 330, 1. (astro-ph/9707167)

Bartelmann M., Huss A., Colberg J.M., Jenkins  A.,  Pearce F.R.,
1997, preprint astro-ph/9709229

Blain A.W. 1998 MNRAS 297, 511 (astro-ph/9801098)

Blanchard A. \& Bartlett J. G. 1997, A \& A 322, L49 (astro-ph/9712078)

Caldwell R. R., Dave R., Steinhardt P. J., 1998a, Phys. Rev. Lett. 80 1582
(astro-ph/9708069)

Caldwell R. R. \& Steinhardt P. J., 1998b, Phys. Rev. D57 6057
(astro-ph/9710062)

Carroll S. M., 1998, preprint astro-ph/9806099

Carroll, S. M., Press, W. H. , Turner, E.L., 1992 Ann. Rev. Astron \&
Astrophys. 30, 499

Casimir H.B.G. 1948 K Ned. Akad. Wet., Proc. Sec. Sci. 51, 793

Chiba T., Sugiyama N., Nakamura T. 1998a, astro-ph/9704199, MNRAS to appear

Chiba T., Sugiyama N., Nakamura T., 1998b,  astro-ph/9806332

Cohen, A., Kaplan, D., Nelson, A., hep-th/9803132

Coble K. , Dodelson S., Frieman J., 1996, Phys. Rev. D55, 1851
(astro-ph/9608122)

Colafrancesco S., Mazzotta P., Vittorio N., 1997, ApJ, 488, 566
(astro-ph/9705167)

Coleman, S., 1988, Nucl. Phys. B307: 867

Davis R.L., 1987, Phys. Rev. D35, 3705

Dienes K. R., 1990a, Phys. Rev. D42, 2004

Dienes K. R., 1990b, Phys. Rev. Lett. 65, 1979

Efstathiou, G., 1995, MNRAS 274 L73

Einstein A. 1917 Sitz. Preuss. Akad. d. Wiss. 1917, 142

Eke, V.R., Cole S., Frenk C.S. \& Henry J.P., 1998 preprint
astro-ph/9802350

Falco E.E., Kochanek C.S. \& Munoz J.A., 1998, ApJ 494, 47
(astro-ph/9707032)

Fan X., Bahcall N.A., Cen R., 1997, Ap J 490 L123
(astro-ph/9709265)

Felten J. E. \& Isaacman R. 1986, Rev. Mod. Phys. 58, 689

Ferreira, P. G. \& Joyce M. 1997a Phys. Rev. Lett. 79, 4740
(astro-ph/9711102)

Ferreira, P. G. \& Joyce M. 1997b preprint astro-ph/9711102

Fukugita M., Futamase T., Kasai M., 1990 MNRAS 246, 24

Fukugita M., Futamase T., Kasai M., Turner E.L. 1992, ApJ 393, 3

Garnavich et al, 1998 preprint, ApJ in press, astro-ph/9806396

Giddings S. \& Strominger A., 1988 Nucl. Phys. B307: 854

Gioia I.M. 1997 preprint astro-ph/9712003

Goldsmith D., 1997, ``Einstein's Greatest Blunder?'', Harvard
University Press

Gott J. R., Park M. G., Lee H. M. 1989, ApJ 338,1

Gross M.A. K., Somerville R.S., Primack J.R., Borgani S.,
         Girardi M., 1997 preprint astro-ph/9711035

Gunn J.E. \& Tinsley B.M. 1975 Nature 257, 454

Guth A. H. 1981 Phys. Rev. D 23, 347

Hancock S., Rocha G., Lasenby A., Gutierrez C.M., 1998, MNRAS 294 L1
(astro-ph/9708254)

Hatano K, Fisher A. \& Branch D., 1997, MNRAS 290, 360

Heath D.J. 1977, MNRAS 179, 351

Henry J. P., 1997, ApJ, 489 L1

Hill C.T. \& Ross G.G. 1988, Nucl. Phys. B 311, 253

Hu W., Eisenstein D.J., Tegmark M., White M., 1998 astro-ph/9806362

Huey G., Wang L., Dave R. , Caldwell R.R., Steinhardt P.J., 1998,
preprint astro-ph/9804285

Kamionkowski M. \& Toumbas N., 1996, Phys.Rev.Lett. 77, 587

Kirshner, R. 1998 talk at CAPP meeting, CERN.

Kochanek C. S., 1993, ApJ, 419,12

Kochanek C. S., 1996, ApJ 466, 638 (astro-ph/9510077)

Kolb E.W. \& Turner M.S., 1990, The Early Universe,
Addison-Wesley Publishing Co.

Lacey C. \& Cole S., 1993, MNRAS 262,627

Lacey C. \& Cole S., 1994, MNRAS, 271, 676 (astro-ph/9402069)

Lahav O., Lilje P. B., Primack J.R., Rees M.J., 1991, MNRAS 251,128

Levin J., 1995, Phys. Rev. D51, 1536 (hep-th/9407101)

Linde A.D., 1982, Phys. Lett. B108, 389

Linde A.D., Linde D., Mezhlumian A. 1995 Phys. Lett. B345, 203
(hep-th/9411111)

Lineweaver C.H., 1998 preprint submitted to ApJL, astro-ph/9805326

Lykken, J., talk at Fermilab missing mass meeting, May 1998.

Maoz D. \& Rix H.-W., 1993, ApJ, 416, 425

Martel H., Shapiro P, Weinberg S., 1998, ApJ 492, 29
(astro-ph/9701099)

Moore, G., 1987, Nucl. Phys. B293, 139

Nieto M. M., 1993, in Gamma Ray-Neutrino Cosmology and Planck Scale
Physics, Ed. D. B. Cline, World Scientific (hep-ph/9212283)

Overduin J.M. \& Cooperstock F.I. 1998 preprint astro-ph/9805260

Padmanabhan T., 1993, Structure Formation in the Universe, Cambridge
          University Press

Peebles P.J.E., 1984 ApJ 284, 439

Peebles P.J.E., 1993, Principles of Physical Cosmology, Princeton
          University Press

Perlmutter S. et al 1995 Ap. J. Lett. 440, L41 (astro-ph/9505023)

Perlmutter S.  et al,1998, Nature 391, 51 (astro-ph/9712212)

Petrosian V., Salpeter E.E., Szekeres P. 1967 ApJ 147, 1222

Phillipps S., 1994, MNRAS 269, 1077

Popowski P. A., Weinberg D.H., Ryden B.S. \& Osmer P.S., 1998
ApJ 498, 11 (astro-ph/9707175)

Press W.H. \& Schechter P., 1974 ApJ, 187, 452

Reichart et al, 1998 preprint astro-ph/9802153

Riess A. G.  et al, 1998 Ap J in press, astro-ph/9805201

Spergel D. N. \& Pen  U.-L., 1997 ApJL 491, L67 (astro-ph/9611198)

Susskind L., 1994  J Math. Phys. 36 6377 (hep-th/9409089)

Tegmark M., Eisenstein D, Hu W. 1998a preprint astro-ph/9804168

Tegmark M., Eisenstein D., Hu W., Kron R., 1998b preprint astro-ph/9805117

Turner E., 1990, ApJ Lett. 365, L43

Turner M.S., Steigman G. \& Krauss L.M., 1984, Phys. Rev. Lett. 52, 2090

Turner M. \& White M., 1997 Phys. Rev. D56, 4439 (astro-ph/9701138)

Van Waerbeke L., Bernardeau F. \& Mellier Y., 1998,
preprint astro-ph/9807007

Viana P.T.P. \& Liddle A.R., 1998, preprint astro-ph/9803244

Vilenkin A., 1984,  Phys. Rev. Lett. 53, 1016

Vilenkin A., 1998, preprint  hep-th/9806185

Wang L. \&  Steinhardt P.J., 1998, preprint astro-ph/9804015

White M. \& Scott D., 1996, Comments on Astrophysics, 19, 289
(astro-ph/9601170)

White M., 1998, preprint astro-ph/9802295

Witten E., 1995, Mod. Phys. Lett. A10, 2153 (hep-th/9506101)

Zlatev I, Wang L, Steinhardt P.J., 1998, preprint astro-ph/9807002
\end{document}